# On the butterfly effect

Alexander Shnirelman
(Concordia University, Montreal, Canada)

The "Butterfly Effect" has been well known and widely discussed in the scientific literature and beyond, especially in the science fiction. It sounds approximately like "The beating of a butterfly wing in South America can result in the considerable change of positions and force of a tropical cyclon in Atlantic 2 weeks later". In other words, the hydrodynamic instability of the atmosphere is so high that small perturbation produced by the motion of a butterfly wing grows exponentially, and the increment is so high that 2 weeks is enough for it to grow to the grand scale. The characteristic time 2 weeks was found by Arnold (see [A], Appendix 2, p.342) as a by-product of his computations of the curvature of the group of volume-preserving diffeomorphisms (in his work the Earth was modeled on a 2-dimensional torus with the size of about 20000km). The curvature computed by Arnold turns out to be negative in most directions, being associated with high instability of geodesics.

Since the work of Arnold, the issue of a strong instability of the flows (i.e. geodesics of the group of diffeomorphisms) attracted considerable attention. In particular, this problem has been studied numerically in [R1, R2]. In this work, the Euler equations on a 2-d torus were solved numerically for several simple initial configurations of vorticity. It was found that there exist several types of behavior similar to scattering. Within each type the solution was moderately stable with respect to small changes in the initial conditions (the perturbations grew linearly), while in the transition zones between different types the solution was highly sensitive to the initial conditions (so that the perturbations grew exponentially in time).

In this work we make a closer look at the flow instability. We solve numerically the Euler equations on the 2-d torus, using a different sort of initial conditions. Namely, the initial vorticity is an isotropic, nearly monochromatic random field (i.e. it is a sum of a finite Fourier series with random coefficients and with frequencies concentrated in a narrow band $|k| \sim k_0$ ). Such initial condition brings about the mechanism of inverse cascade which results in merging of small vortices and formation of larger ones. The final configuration is a pair of large vortices of opposite signs. These vortices slowly move about; their shapes are slowly pulsating. In some simulations, we can observe smaller "satellites" of larger vortices. The flow (which is a result of a long evolution) appears to be quasi-periodic in time.

In order to reveal the possible strong instability, we ran two series of numerical simulations with very close initial conditions. Namely, the changes were made in the amplitude of just one mode; the relative difference was about $10^{-7}$.

The flow picture in both solutions at different moments is shown in the following series of pictures. In each of the pictures, the left one corresponds to the first solution, and the right picture corresponds to the second one. We can see that two solutions are practically indistinguishable at the beginning; however, they diverge later, and soon look as completely unrelated to one another. Their final configurations are just the pairs of large vortices; but these vortices are at very different positions. So, if we regard the initial perturbation as produced by a "butterfly", it results in a considerable change of position of large vortices ("cyclons") a while later. So, the "butterfly effect" is materialized literally.

The perturbation growth is shown at Figure 7. We see that the perturbation amplitude, measured in the vorticity norm, grows from about $2 \cdot 10^{-8}$ to about $10^0$.

The simulations were done using the standard pseudospectral method. The resolution was $2^{10} \times 2^{10}$ modes; a small viscosity $\eta = 0.5 \cdot 10^{-5}$ was added to ensure the numerical stability. The characteristic wave number of the initial flow $k_0 \sim 20$, so that the initial configuration consisted of about 400 vorticity blobs. The simulation results are quite robust, and 100% reproducible.

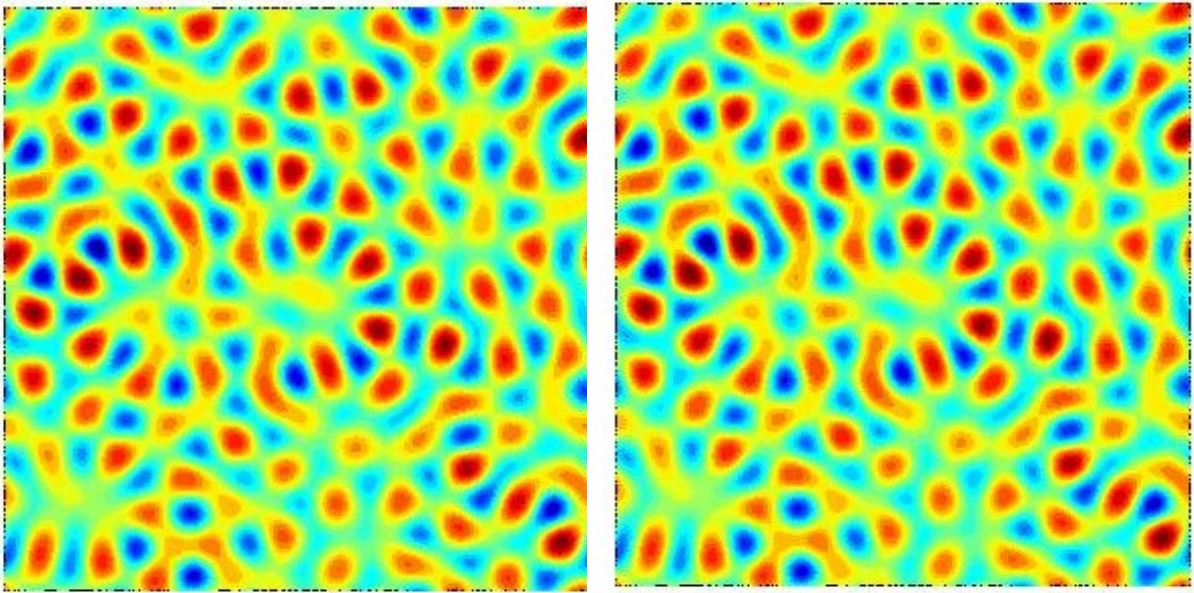

Figure 1. Initial vorticity distribution at $t=0$. The left and the right pictures here and in the next figures correspond to two simulations with very close initial conditions.

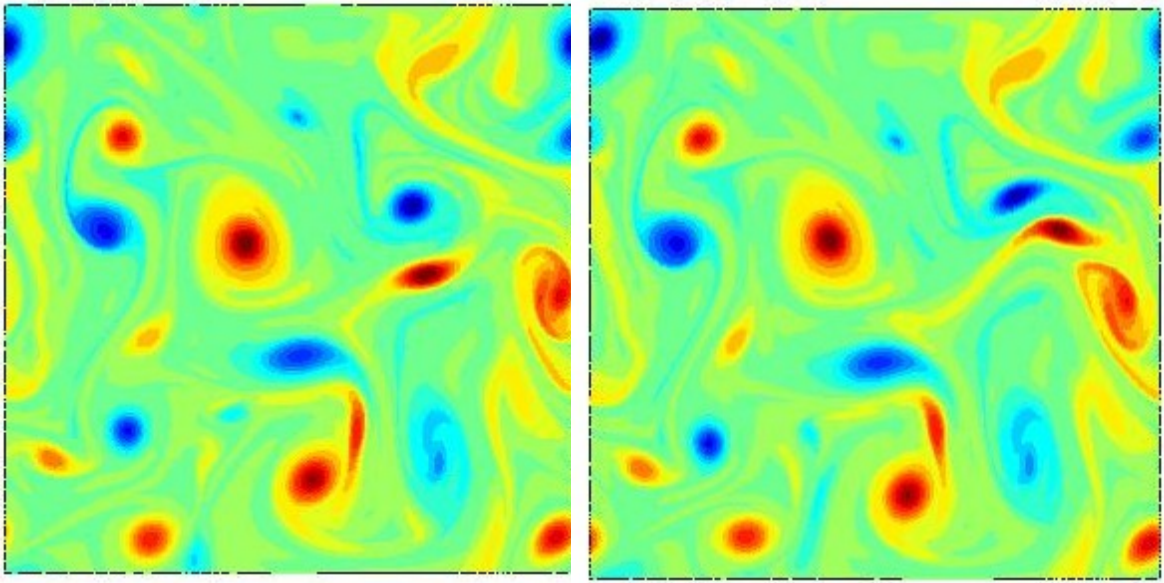

Figure 2. Vorticity distribution at $t=20$.

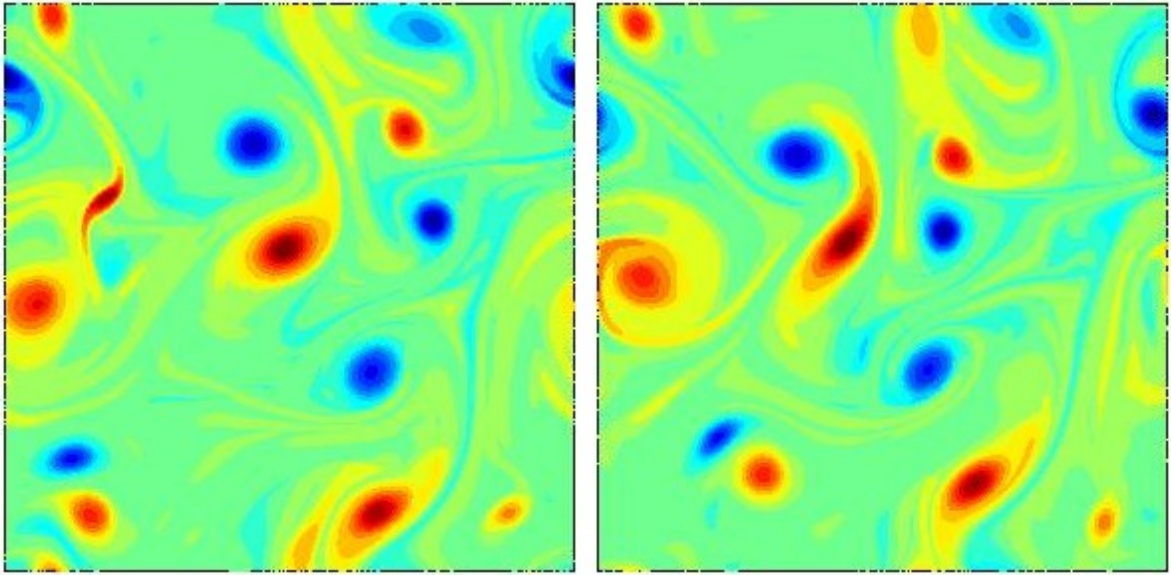

Figure 3. Vorticity distribution at $t=25$.

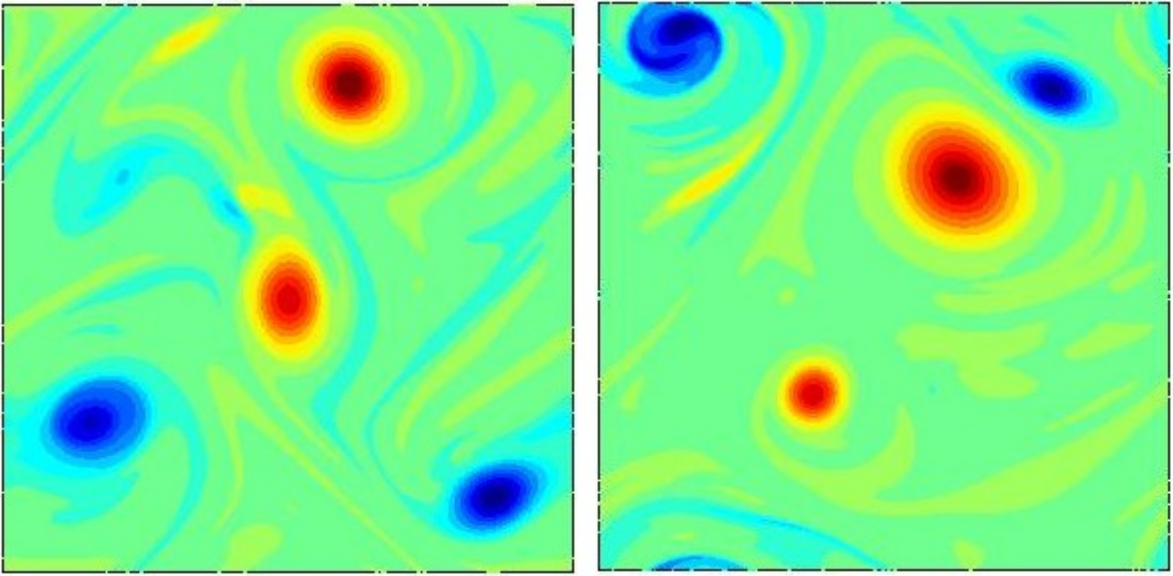

Figure 4. Vorticity distribution at $t=60$.

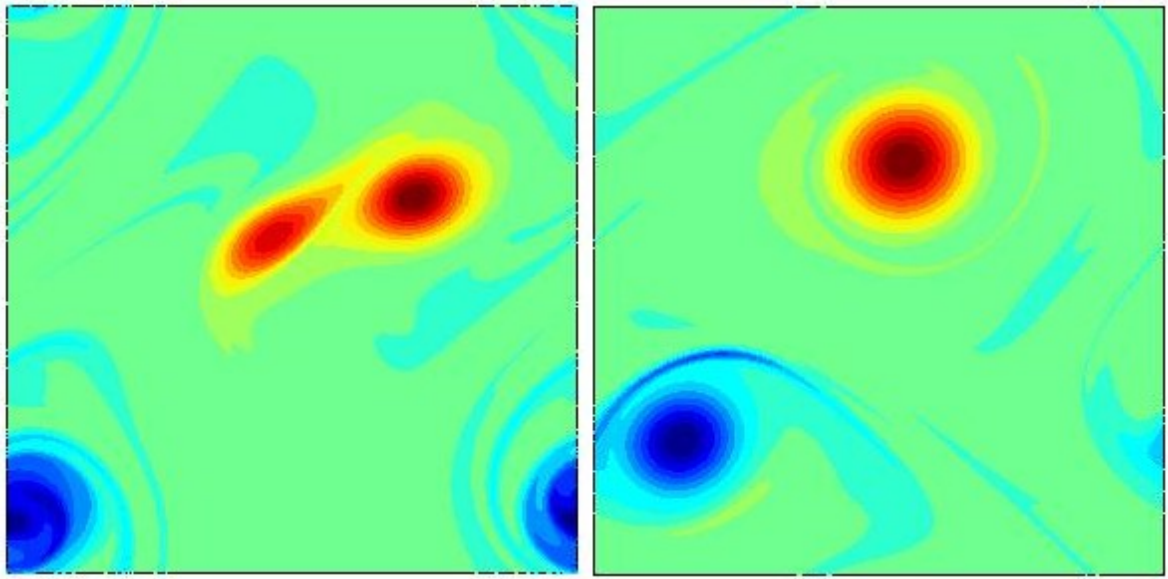

Figure 5. Vorticity distribution at $t=100$.

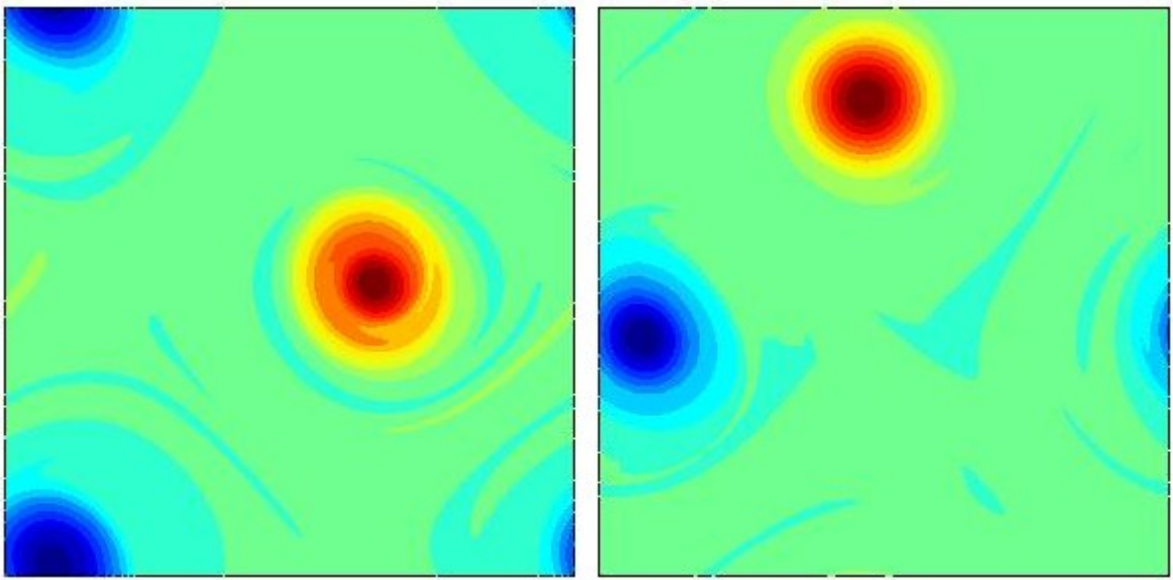

Figure 6. Vorticity distribution at $t=120$.

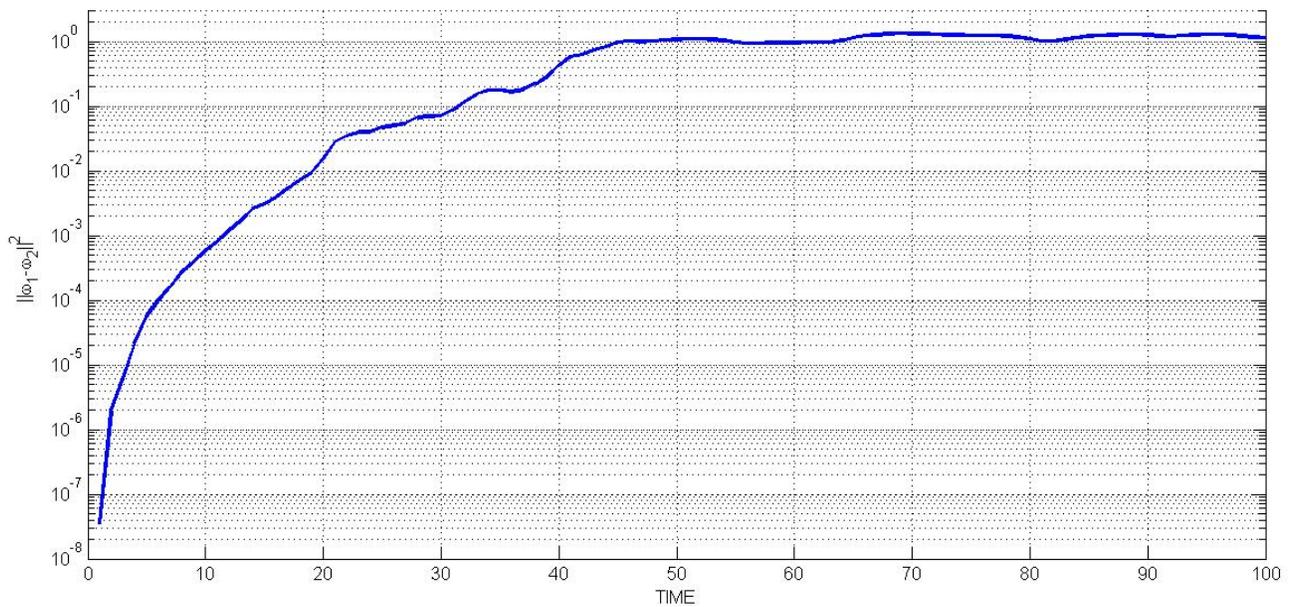

Figure 7.  $L^2$ - norm of the perturbation as a function of time.

So, the "Butterfly Effect" exists.

## REFERENCES

[A] V.Arnold, Mathematical methods of classical mechanics. Springer, 1989.

[R1] Raoul Robert, L'effet Papillon n'existe plus! SMF – Gazette – 90, Octobre 2001 .

[R2] Raoul Robert, Les prévisions météorologiques seraient impossibles au-delà de deux semaines, à cause de l'« effet papillon ».
https://interstices.info/jcms/c_19155/l-effet-papillon-n-existe-plus